# Machine learning approaches in Detecting the Depression from Resting-state Electroencephalogram (EEG): A Review Study

Milena Čukić Radenković[1,2], PhD

[1]Department for General Physiology and Biophysics, University of Belgrade, Belgrade, Serbia

[2]Amsterdam Health and Technology Institute, HealthInc, Amsterdam, the Netherlands

**Abstract**

In this paper we aimed at reviewing several different approaches present today in search for more accurate diagnostic and treatment management in mental healthcare. Our focus is on mood disorders, and in particular on major depressive disorder (MDD). We are reviewing and discussing findings based on neuroimaging studies (MRI and fMRI) first to get the impression of the body of knowledge about the anatomical and functional differences in depression. Then, we are focusing on less expensive data-driven approach, applicable for everyday clincal practice, in particular those based on electroencephalographic (EEG) recordings. Among those studies utilizing EEG, we are discussing a group of applications used for detecting of depression based on the resting state EEG (detection studies) and interventional studies (using stimulus in their protocols or aiming to predict the outcome of therapy). We conclude with a discussion and review of guidelines to improve the reliability of developed models that could serve improvement of diagnostic of depression in psychiatry.

Introduction

Current practice in mental health institutions is in crisis. The depression is very high on the list of psychiatric disorders causing the long sick leaves, and yet the diagnostic for that particular disorder is still relying on a conversation only. Current clinical practices do not use in their diagnostic process any of physiological or biochemical tests to confirm, for example, the presence of certain biomarkers in a person's body, before prescription of the medication. In their study about psychiatry research online, other authors commented that 'unlike the rest of the medicine, psychiatry has no objective diagnostic tests' (Gillan and Daw, 2016). Shorter and Fink are in line with them, as well as the whole list of respectable psychiatrists asking for the change in how DSM is defining diagnosis for depressive disorders (Parker, Fink, et al., 2010) without any biochemical biomarker included. They stated that 'melancholia's features cluster with greater consistency than the broad heterogeneity of the disorders and conditions included in major depression and bipolar disorder.' The importance of a medically trained interviewer's role to collect the data (in a clinical setting, psychiatric interview for diagnosis of depression) from patients became controversial, for it is shown that clinicians-rated measures are in critical aspects less reliable than a self-report taken from patients anonymously and directly in an online collecting of information (Berinsky et al., 2012). Also, the number of patients requiring psychiatric attention is on the rise. Among more than 300 milions of people suffering from some depressive disorder, just twenty percent of them are receiving professional help (Whitefold et al., 2013; World Economic Forum, 2019). Experts for financial side of healthcare calculated that the depression is going to reach enormous $16 trilion dollars costs before 2030, and WHO published its prediction that before the very same year, 2030, it will become world health issue number one (Mathews and Loncar, 2006). Certainly, there is room for improvement of how we define, detect and treat mental health problems, especially depression which is the top cause of sick days leave and the years spent with disability (DALY, 2015; WHO, 2017). According to Professor Damian Denys, Head of the Netherlands Association of Psychotherapists (September 2018, NRC Interview) mentioned that the waiting lists to visit a psychiatrist are between six and nine months. For example, in the Netherlands, 75% people are asking 'Dr. Google' for their symptoms, instead of visiting the doctor (the Netherlands is the leader in that trend in Europe in 2018). We will try in this text to mention, describe and discuss different Online tools, research going online, comlexity measures used in electrophysiology, Data mining and machine learning , as well as various combinations of those tools and methodologies possible with technological innovations we have today.

**How online and mobile applications can serve detection of depression?**

Today, everything is online. If someone feels sick, the probability that she will search her symptoms online, before contacting the doctor, is high. Everyone is accustomed to that. Advances in technology made available wearables and smart whathces, mobile applications, the devices monitoring the quality of sleep and also collecting the daily reports from the outpatients. That kind of information could be valuable for the mental healthcare system in the sense that those who get the indication to seek help are encouraged, and this is already helping the classical way of triage in primary care. In their paper 'Taking psychiatry research online', Gillan and Daw are explaining the benefits of this approach. Those who are already socially isolated, disabled or unable to commute, can reach the help online (Gillan and Daw, 2016). There are several organizations offering online counseling to people who otherwise cannot reach professional help (for example, DeSofa). A research in Psychiatry can be organized and designed as usual , but rely on online surveys, and therefore, include those who are ussualy 'invisible' to healthcare system.  Besides, both Google and Amazon have their online application in connection with depression; on Google you can ask the question 'am I depressed?' and get the access to the standard questionnaires to inform yourself whether you would need to seek the help. Amazon started in 2008 a 'citizen scientist' initiative, with the Mechanical Turk platform as a means of participation and a trial to have clearer picture how many people who do not report their depression are out there. Also, there are publications already testing this approach to collecting the data, and confirming that it is valuable, especially because of the stigma still present in our societies and the confortable anonimiety during the process (Berinsky et al., 2012; Suhara et al., 2017). Data analytics imposed an additional problem to policy makers (a societal value of data in context of urban regions), but this possible aggregation of data on larger scale can work only if fellow citizens contribute to the aggregated data. Suhara and colleagues published DeepMood: Forecasting Depressed Mood Based on Self-Reported Histories via Recurrent Neural Networks, in 2017. They collected large-scale records from 2382 self-declared depressed people and analyzed their data. Recurrent network algorithm incorporates categorical embedding layers for forecasting depression. The experiment showed that their method can forecast the severely depressed mood of a user based on self-reported histories, with higher accuracy than support vector machines (SVM). Their results also showed that the long-term historical information of a user improves the accuracy of forecasting depressed mood (Suhara, Xu and Pentland, 2017).

**What is Computational Psychiatry?**

When computational neuroscience meets psychiatry an utterly new field emerges called computational psychiatry (Montague et al., 2011; Wang and Krystal, 2014; Yahata et al., 2017). With the advent of very sophisticated methods of neuroimaging, different findings emerged which might serve psychiatry to understand the (micro) biological and physico-chemical underpinnings of mental disorders. Those possible causes can vary from molecular to the cell to synaptic and neural networks level. There is more and more ongoing research directed to understanding the mechanisms characteristic for depression. There are two groups: a theory-driven and data-driven research. Major contributors to that body of evidence are coming from different neuroimaging techniques and methods to mention just the major ones: functional magnetic resonance imaging (fMRI), MRI, Digital tractography imaging (DTI), Functional anisotropy (FA), Magnetoencephalography (MEG), Electroencephalography (EEG), etc. There are quite a few successful researches using fMRI data to run machine learning in an attempt to find the biomarkers (or neuromarkers) for depression (for example, see Chen et al., 2015). The data-driven approach typically involves machine learning. There are a rising number of researchers working on such a task (for review see Yahata et al., 2017) repeatedly showing that certain features can be used for the detection or for elucidating of missing links important for understanding this severe disease. If you read online news and reports it might seem that the future of personalized medicine is already here; which is not the case (Stephen et al, 2017). One of the reasons why those methodologies are still not utilized in everyday clinical practice is their high price and the second reason is that there are a lot to be done for generalization to be ready for daily clinical use. The training phase still has to be performed each and every time, and the accuracy on the test phase is varying quite a lot, which is too much for medical approval of the possible application. Technical reasons why additional work has to be done before translation to mental healthcare is the issue of overfitting and the disproportion of number of variables and the size of the sample and we need improvement to the level of thousands of participants. A possible solution could be to organize a multicenter collection of data among a big number of institutions on a joint project (like RDoC Project, STAR*D, IMAGEN) or start co-recording EEG with every MRI, so much more affordable research can rely on that data.

**Some of the findings from neuroimmaging research tackling the biomarkers for detection of depression**

We are going to mention some of the main findings here, notable for one to understand the extent of present knowledge about the possible neuromarkers for depression. Bluhm examined the resting state default-mode network connectivity in early depression using a seed region of interest analysis (Bluhm et al., 2009). They confirmed the decreased connectivity within the caudate nucleus. Direct comparison showed significantly reduced correlation between precuneus/posterior cingulate cortex and the bilateral caudate in depression compared with controls and no areas of interest connectivity in the depressed group. Vederine et al. (2011) elaborated on abnormal functional connectivity in the fronto-limbic system, as well as de Kwaasteniet (Kwaasteniet et al., 2013). De Kwaasteniet confirmed by the combination of fMRI and FA that second part of uncinate fasciculus, which is the deep white matter tract connecting prefrontal with the limbic system, is not completely functional. The cause for this deficit is not known; it might be previous trauma or too long exposure to cortisol, or something else. Many other studies confirmed the abnormal functional connectivity, but we will mention here those who used graph theory analysis and machine learning to use previous knowledge of different markers to detect depression. Zhang and his colleagues (2011) published fMRI/graph theory (small world) study confirming disrupted brain connectivity networks in drug-naïve first-episode major depressive disorder (MDD). It seems that MDD disrupts the global topological organization of the whole-brain networks. Zeng and colleagues used unsupervised classification of major depression based on functional connectivity MRI (Zeng et al., 2013). With the group-level clustering consistency of 92.5% and individual-level classification consistency of 92.5%, they extracted sufficient information from the subgenuate cingulate functional connectivity map to differentiate depressed patients from controls. They showed that subgenuate cingulate functional connectivity network signatures may provide a promising objective biomarker for detection of depression (MDD). Ramasubbu and colleagues tested the accuracy of automated classification of major depressive disorder as a function of symptom severity (Rajamannar Ramasubbu et al., 2016). The resting state fMRI data showed statistically significant accuracy only for the very severe depression group (66% accuracy), while mild to moderate and severe depression were just little higher than a chance to detect (consequently 58% and 52%). Binary linear SVM classifiers achieved a significant classification of very severe depression only. Guo and colleagues examined multi-feature approach based on a high-order minimum spanning tree functional brain network (Guo et al., 2017) to classify patients with MDD. They obtained classification accuracy of 97.54%. One of the most exciting results published by Tokuda and colleagues (Tokuda et al., 2018) identified three subtypes of depression based on fMRI scans, patient histories and formally collected data. They applied statistical learning for discovering the subtypes. For this study data about clinical, biological and life history from 134 individuals were collected. MRI was used to map brain activity patterns in different regions; they examined 78 different regions covering the entire brain. This is the first study to identify depression sub-types from life history and MRI data. With over 3000 features, including whether or not participants had experienced childhood trauma, the team developed a novel statistical tool to extract relevant information for clustering similar subjects (multiple co-clustering methods). The focus of the study was to detect various ways of data clustering and the features responsible for it. Three out of five data clusters were found to represent different sub-types of depression. All three were characterized by two main factors: functional connectivity patterns synchronized between different regions of brain and childhood trauma experience. The brain's functional connectivity in regions that involved the angular gyrus ( associated with processing language and numbers, spatial cognition, attention and other aspects of cognition) which played an important role in determining whether SSRIs were effective in treating depression. Participants who had increased functional connectivity and who also had experienced childhood trauma had a sub-type of depression that is unresponsive to treatment by SSRIs drugs (D1). Two other identified types which tended to respond positively to treatments using SSRIs drugs did not show increased connectivity (D3) or did not experience childhood trauma (D2). This finding is about neurobiological aspects of depression not described before; they could help psychiatrists and therapists to improve the accuracy of their diagnostic and treat their patients more effectively (Tomoki Tokuda et al., 2018).

There are also other avenues of research trying to understand mental disorders as aberrations of normal functioning recognized via disrupted brain connectivity (van Essen et al., 2012; Castellanos et al., 2013; Kim et al., 2013). All those lines of research are aiming at extracting different biomarkers, or neuromarkers for specific mental illnesses; we are going to focus here on depressive disorder only, although certain overlapping and similarities in mechanisms can be observed with other psychiatric disorders (Li et al., 2008).

One of the seminal works for the application of graph theory (the base for connectomics) in neuroscience, by Olaf Sporns, is 'Networks of the Brain' (2011). He pointed out that the forefather of the whole theory behind today's connectomics was actually Norbert Wiener, who stated in his 'Cybernetics' some of the basic ideas of control and communication theory (Wiener, 1948). In the chapter named 'Cybernetics and Psychopathology' he stated that '(there is) …*nothing surprising in considering the functional mental disorders as fundamentally diseases of the memory, of the circulating information kept by the brain in the active state, and of the long-time permeability of synapses*' which is in line with recent recurrent dynamics and functional integration of the brain. Wiener discussed the cases of several mental illnesses such as schizophrenia, depression, and paranoia which he called 'functional mental disorders' (Wiener, 1948). It turned out that he was right because several recent findings of the mechanisms underlying depression (Kwaasteniet et al., 2013) and bipolar depression (Kim et al., 2013) confirmed exactly the same thing: disrupted functional connectivity within fronto-limbic system important for depression.

**Comeback of EEG for detection : Re-use of an existing equipment**

Another solution to the problem of detection of depression might be focusing our attention to well established and noninvasive electroencephalogram (EEG) which is considered to be the oldest neuroimaging technique. Many researchers explore this approach due to its accessibility to a big number of patients and cost-effectiveness. Another advantage of EEG is its presence in almost every modern hospital, so basically we are proposing here re-use of already existing equipment. What is required here is just another mode of application in analysis of the recorded signal. Until now EEG was used in psychiatry for confirming the epileptic forms only. In the last eight years we witnessed the burst of research in this field, with different aims though. Some were done as an attempt to prove that the detection is possible (Ahmadlou et al., 2012), some compared classical spectral markers with nonlinear ones (Hosseinifard et al., 2014), and some tried to show how high accuracy can be improved (Faust et al., 2015). Our previous work demonstrated (Čukić et al., 2018) that if one characterize the signal (EEG) with complexity measures, any of seven most popular classifiers is resolving the task with high accuracy (between 87% and 97%). Although it is shown that in electrophysiology classical spectral measures become redundant with fractal and other nonlinear measures (dynamic symbolic approach, geometrical techniques, statistical measures, information measures, entropy measures, and fractal analysis) many clinicians still rely on them. Also, the broadband analysis showed to be much more successful than dividing signal on bands, which never demonstrated its physiological purpose (Baçar eta l., 2011). Another line of research is developing wireless EEG caps (Epoch, ENOBIO Neuroelectrics, iMotions, just to mention the few) which can be used for research in the environment without restraining the patient, and even for monitoring of recovering from severe episodes. If wireless EEG recorder would become accessible soon, we are sure that early detection and timely intervention will prevail fast.

**Resting state EEG research in Depression Detecting task**

In the present literature, there are several approaches in examining the changes in complexity of EEG characteristic for depression. Among researchers there seem to be appearing a consensus that the characteristic of depression is elevated complexity of EEG when compared to healthy peers (for review see, de la Torre-Luque, 2017). From previously published fMRI and DTI (FA) studies (de Kwaasterniet et al., 2013; Vederine et al., 2011) and graph theory applications on EEG signal (Kim et al., 2013) we know that in different depressive disorders changes in functional connectivity are confirmed. Deep white-matter tracts important in the fronto-limbic system seem to be damaged (the second part of uncinate fasciculus, in MDD) probably introducing further functional changes in the very complex system of interactions. That is possibly reflected on the excitability of cortex so we could detect the difference in EEG between people diagnosed with depression and healthy controls (Ahmadlou et al., 2011; Bachmann et al., 2012; Faust et al., 2014). The conclusion of de la Torre-Luque reviewing study (2017) demonstrated very different approaches showing elevated complexity in the majority of published papers, concluding that 'EEG dynamics for depressive patients appear more random than dynamics of healthy non-depressed individuals.' Also, there is a consensus about utilization of more than one nonlinear measure, because different measures are detecting unique features of the EEG signals 'revealing information which other measures were unable to detect' (Burns and Rayan, 2015).

Based on that field of research, further applications followed. Data mining is reportedly 'the extraction of implicit, previously unknown, and potentially useful information from data' (Witten and Frank, 2005), and machine learning as a part of that discipline attracted a lot of attention lately. From period 2012 to 2018 twelve studies reported about the successful classification of depression based on nonlinear features calculated from resting-state EEG, including our own work. The problem with this cohort of studies is similar like with those trying to elucidate the changes of complexity from depressed patients EEG; the direct comparison is challenging due to very different methodologies used. This work aims at comparing those different approaches to give a glimpse of the future capacity of using this approach in everyday clinical practice.

One of the first studies using resting state EEG to classify depressed persons and healthy controls were the one by Ahmadlou et al., 2012. They also aimed at comparing two different algorithms for calculating the fractal dimension. What they found is in line with work of several other authors (Esteller et al., 1999 and 2001; Castiglioni, 2010).

Ahmadlou and colleagues wanted to examine which of two algorithms for calculating the fractal dimension (FD) is better as a feature of resting state EEG for classification of MDD from healthy adults: Katz FD or Higuchi FD (HFD). Based on Katz FD (KFD) algorithm they previously had successful study in finding a biomarker for Alzheimer's disease (AD; Ahmadlou et al., 2011b). From previous literature, they knew that different algorithms for calculating FD may differently interpret self-similarity and irregularity of time series since they are calculating it in different ways. Esteller et al., (2001) showed that KFD is more robust to noise compared to Petrosian's and Higuchi's algorithm, but other showed (Raghavendra and Narayana 2009; Castiglioni, 2010) that KFD is dependent on sampling frequency, amplitude and waveforms, which is a disadvantage in analyzing biosignals. The disadvantage of HFD is, according to Esteller (2001) that it is more sensitive to noise than KFD. The sample for this study comprised young adults (20 to 28 years), 12 non-medicated MDD patients and 12 healthy controls (both groups had 5 women and 7 men). The DSM-IV and Beck depression scale was used for scores for MDD. In their experiment resting state EEG was recorded for 3 minutes, with closed eyes, and sampling rate was 256Hz. They opted to record just frontal electrodes for this experiment (7 electrodes Fp1, Fp2, Fz, F3, F4, F7 and F8, 10/20 standard) because of the previous finding that prefrontal cortices showed abnormal functioning in MDD. They used wavelet (Daubechies wavelet of order 44, 4-4level wavelet decomposition) to decompose the raw EEG signal into 5 standard sub-bands (gamma, beta, alpha, theta, and delta). Klonowski showed (2007) that wavelet is a little bit better than classical Fourier's decomposition, but still distorting signal under study. Ahmadlou and team used averaged calculated KFD and HFD values (they divided it on the left and right electrodes and averaged it) as features for Enhanced Probabilistic Neural Networks (EPNN) after application of ANOVA which evaluated the ability of a feature to discriminate the groups based o variations both between and within groups. EPNN had an input layer consisting of nodes equal to a number of selected features by ANOVA, and this robust classifier insensitive to noise in training phase used Bayesian rules. Output layers had nodes for each class (Depression or Healthy). They found that MDD and non-MDD are more separable in the beta band based on HFD (contrary to previous belief that the differentiation is best in alpha band) and that HFD in both beta and gamma bands in MDD is higher than in healthy participants. That implied on higher complexity of signal recorded from frontal cortices (according to their data left frontal lobe is more affected). Based on HFD (which performed better than KFD) they obtained high accuracy of 91.3%.

After Ahmadlou and his colleagues, followed several studies by Indian scientists.

Subha Puthankattil and her colleagues (2012) aimed at classification of EEG records of depressed patients and controls by utilization of relative wavelet energy (RWE) and artificial feedforward neural network. Their sample comprised of 30 depression patients (aged 20-50 years) and the age and gender-matched healthy controls (16 females, 14 males). There is no information on whether the patients were medicated or not.

The recording was taken from four locations in total; FP1-T3 (left) and FP2-T4 (right hemisphere). For 5 minutes they recorded resting state EEG (the information about closed or open eyes is missing), sampling frequency was 256Hz, and they used notch-filter, but also utilized total variation filtering (TVF) for high-frequency noise. (They used a similar method in previous research to classify epilepsy; (RWE details and approximations used to select a suitable wavelet for artifact removal in the EEG, plus linear discriminant analysis- LDA and SVM). The signal was divided into sub-bands, eight levels multiresolution decomposition method of discrete wavelet transform (DWT) was used. RWE analysis (feature extraction) provided information about the signal energy distribution at different decomposition levels; twelve features were extracted for training and testing NNs. Nine features included values of RWE for different frequency bands, and two were obtained by observing the trend of the variation of the average RWE of EEG signals. The signal energy RWE is higher in depression. Coif let 5 showed the highest correlation coefficient and indicated the best match for depression patients EEG (out of an array of 23). The performance of artificial neural networks yielded an accuracy of 98.11% (normal and depression signals). Sensitivity 98.7%; Selectivity 97.5%; specificity 97.5%.

Hosseinifard and colleagues (2013) examined nonlinear analysis of EEG in 45 unmedicated depressed patients (23 females; 20-50 years old) and 45 healthy controls (19 to 60 years old). DSM-IV interview and Beck Depression Inventory were used. Their study aimed to classify healthy and depressed persons and to improve the accuracy of classification. The resting state EEG with eyes closed was recorded for 5minutes from 19 electrodes from 10/20 standard system. Sampling frequency was 256Hz, they used high and low plus notch filtering. Artifacts were discarded after visual inspection. They divided raw EEG signal into standard sub-bands and applied both classical (Welch method, a power of four EEG bands) and four different nonlinear measures (detrended fluctuation analysis (DFA) Higuchi fractal dimension (HFD), correlation dimension and Lyapunov exponent) to characterize the signal. After feature extraction was performed, one classical and four nonlinear measures were calculated for all 19 electrodes for each person. Each feature vector (19 features related to 19 electrodes) is applied to K-nearest neighbors (KNN), Linear Discriminant Analysis (LDA) and Linear Regression (LR) classifiers. Two third of the sample was used for the training phase and the remainder for the test set. Leave-one-out cross-validation (LOOCV) method was applied to the training dataset and genetic algorithm for feature selection. Based on selected features of the training phase, the testing phase is performed. For feature selection they used a genetic algorithm (GA) the size of the population is set to 50, cross over rate to 80% to %5 (They also tried PCA, but GA outperformed it significantly). Classification accuracy was the best in the alpha band for LDA and LR both reaching 73.3% (the worst was KNN in delta and beta and LDA in the delta with 66.6%). The best accuracy in the experiment was obtained by LR and LDA classifiers. The accuracy of all classifiers increased when the signal was characterized with nonlinear features, nor classical power (LR reached 90% with correlation dimension). The conclusion was that 'nonlinear features give much better results in the classification of depressed patients and normal subjects' contrary to classical one. Also, they concluded that depression patients and controls differ in the alpha band more than other bands, especially in the left hemisphere.

Faust and colleagues (2014) used EEG signals in the automated detection of depression. They probably used the same sample as Subha D. Puthankattil, and Paul K. Joseph, namely 30 patients diagnosed with clinical depression and 30 controls (age from 20 to 50; 16 females and 14 males) recorded earlier (2011/12). Only 4 electrode positions were used, left FP1-T3 and right FP2-T4. They also used wavelet packet decomposition (WPD, Db8 wavelet) to extract appropriate sub-bands from the raw signal (feature extraction). Those extracted sub-bands were input for calculating several entropy measures; Bispectral entropy (Ph, including Higher-order spectra HOS technique, from Fourier analysis), Renyi entropy (REN), Approximate entropy (AppEn) and Sample entropy (SampEn). The process of extraction the sub-bands comprised of sending the original data through a sequence of down-sampling and low pass filters which defined the transfer function (which is like classical spectra analysis distorting the information content of the data, according to Klonowski, 2007). Also, prior to that extraction, researchers claim that high-frequency components did not contribute relevant information (contrary to our findings, Čukić et al., 2018) and removed them as well. After applying Student's t-test to evaluate features, several classification algorithms were applied. The aim was to extract the important information in order to differentiate individual categories; Gaussian Mixture Model (GMM), Decision Trees (DT), K-nearest neighbors (KNN), Naïve Bayes Classifier (NBC), Probabilistic Neural Networks (PNN), Fuzzy Sugeno Classifier (FSC) and Support Vector Machines (SVM). In this experiment, they applied ten-fold stratified cross-validation. The accuracy was 99.5%, sensitivity 99.2%, and specificity 99.7%. Contrary to Hosseinifard they claim that the EEG signals from the right part of the brain discriminate better the depressive persons.

The sample used for another study (Acharya et al., 2015) comprised of 15 healthy and 15 depressed persons (20-50-year-olds).(Again recorded in Medical College Department of Psychiatry Calcut, Kerala, India, possibly the same sample like in Sudha D. Puthankati 2012 and Oliver Faust et al. 2014, the details are exactly the same). Resting state EEG was recorded from just four positions in 10/20 standard system, left FP1-T3 and right FP2-T4, during 5 minutes (eyes closed and eyes opened). The sampling rate was 256Hz, notch-filtered (also comprising from 2000 samples like in Faust and Puthankattil), artifacts were manually removed. Feature extracting consisted of 15 different measures: fractal dimension (Higuchi fractal dimension, HFD), largest Lyapunov exponent (LLE), sample entropy (SampEn), DFA, Hurst's exponent (H), higher order spectra features (weighted Centre of bispectrum, W_Bx, W_By), bispectrum phase entropy (EntPh), normalized bispectral entropy (Ent1) and normalized bispectra squared entropies (Ent2, Ent3), and recurrence quantification analysis parameters (determinism (DET), entropy (ENTR), laminarity (LAM) and recurrent times (T2)). These extracted features are ranked using the t value. The information whether some of them were calculated on standard EEG sub-bands and other on broadband signal is missing (like classical spectral measure high order spectra utilizing Fourier's analysis, must have been computed in sub-bands, but that was not mentioned). After a large number of trials, the authors decided based on a comparison of values to formulate Depression Diagnosis Index taking into account only LAM, W_By and SampEn, without the explanation. It says that 'DDI is a unique formula that yields nonoverlapping ranges for normal and depression classes.'

So, that index is depending only on three nonlinear features. This (probably) heuristically obtained index is used here instead of usually utilized classifiers. Features are ranked based on t value and fed to classifiers one by one obtaining the accuracy higher than 98%, sensitivity higher than 97% and specificity more than 98.5%. This best result is reportedly obtained by utilization of SVM with a polynomial kernel of order 3 (for both left and right hemisphere; they used averaged values for left and right hemisphere), although in previous papers by the same authors SVM as discarded before. The text has an ambiguity in 'features are fed to SVM classifier' and in the next sentence 'SVM classifier yielded the highest classification performance with the average accuracy…' Whether SVM is actually used is among other inconsistencies in this groups' work report.

Contrary to Faust, who analyzed the same sample, here SampEn among other features is showing significantly higher values for depression than controls

Bairy et al. (2015) used a discrete cosine transform (DCT) to decompose the raw EEG (of depressed and healthy persons) data to frequency sub-bands. Further, they calculated sample entropy, correlation dimension, fractal dimension, Lyapunov exponent, Hurst exponent and detrended fluctuation analysis (DFA) on DCT coefficients and the characteristics features are ranked by utilization of t-value. These features are used as input for classifiers DT, SVM, KNN, and NB. SVM with radial basis function (RBF) yielded an accuracy of 93.8%, the sensitivity of 92% and specificity of 95.9%. We cannot say whether internal or external validation was performed, nor the details of, for example, the method used to calculate fractal dimension, hence the reproducibility of this study which claims so high accuracy is close to zero.

Another study is published in the same year; Mohammadi et al., 2015. This study used a sample of 53 MDD patients and 43 HC.EEG recordings lasted for 3 minutes (vigilance controlled eyes closed and eyes opened) in resting condition. The sampling rate was 500 Hz, a bandpass filter was used. Brain Vision Analyzer Software, automatic artifact rejection. 28 electrodes are used for analysis, FFT, Fourier transform for calculating power for standard sub-bands (classical spectral measures were used) a total of 12 datasets existed, min-max normalization. After preprocessing the data, cleaning, and normalization, they applied Linear Discriminant Analysis (LDA) to map features into a new feature space (data evaluation phase). They applied Genetic Algorithm (GA) to identify the most significant features. They build predictive models with Decision Tree (DT). From two experiments, the first analyzed each frequency ban individually, while the second experiment analyzed the bands together. The model showed an average accuracy of 80% (MDD vs. HC). When one compares their methodology with previously mentioned studies, its layers are so complicated and strenuous that it is hard to believe that clinicians would apply it. There is no clear information about the checking of liability of their high accuracy nor both internal and external validation (in terms of good generalization).

Liao et al. (2017) proposed a method based on scalp EEG and robust spectral-spatial EEG feature extraction based on kernel eigen-filter-bank common spatial pattern (KEFB-CSP). They first filter the multi-channel EEG signals (30 electrodes traces used in this experiment) of each sub-band from the original sensor space to new space where the new signals (i.e., CSPs) are optimal for the classification between MDD and healthy controls, and finally applies the kernel principal component analysis (kernel PCA) to transform the vector containing the CSPs from all frequency sub-bands to a lower-dimensional feature vector called KEFB-CSP (with 80% accuracy). Their sample comprises of 12 (4 males, 8 females, 43 -70 years old) patients plus 12 healthy controls, and resting state EEG was measured for 5 minutes. Comparison with different classifiers in the single-trial analysis: KNN, LDA, and SVM. They also used the majority voting strategy based

Mumtaz and colleagues published their study on the same year as Liao: 2017. It relied on resting state EEG, 33 MDD patients, and 30 Healthy controls. Measures were spectral power of different frequency bands for diagnosing depression. Noise removal and feature extraction. Matrix was formed from extracted features, z-score standardization, according to its mean and variance. To determine the most significant features a weight was assigned to each feature based on its ability to separate the target classes according to the criterion, (ROC). Only the most significant features were used for training and testing the classifier models: Logistic regression (LR), SVM and Naïve Bayesian (NB). The models are validated with application fo 10-fold cross-validation that has provided the metric for accuracy sensitivity and specificity. LR (acc 97.6%, sens 96.66%, spec 98.5%), NB (acc 96.8%, sens 96.6%, spec 97.02%), SVM (accuracy 98.4%, sensitivity 96.6% and specificity 100%).

Yet another study utilizing resting state EEG was published by Bachman et al., 2018. In this study (after the first one from 2013 when they explore the possibility of detection with novel spectral index SASI and Higuchi FD, but back then with k=50), Bachmann and colleagues performed classification task with resting state EEG measured from just one electrode. Their sample comprises of 13 depression patients (medication-free) and 13 healthy controls; they analyzed alfa power variability and relative gamma power, but also Higuchi's fractal dimension and Lempel-Ziv complexity. From 30 channel EEG, they opted on relying on a record from just one electrode. Features were used for classification by utilization of logistic regression with leave-one-out cross-validation. They reached maximal accuracy of 85% with HFD and DFA, but also HFD and LZC, and for only one nonlinear measure maximal 77% (much lower than the performance of our average classifiers). This time, for calculating HFD they used k=8 (like in all of our studies so far). Researchers claim that accurate detection of depression is possible with utilization the record from only one electrode and explain that with the eventual acceptance from clinicians who prefer straightforward methods of detection. They also concluded that' there is no single superior measure for detection of depression.'

In our study from 2018 (Čukić et al. 2018), we aimed to elucidate the effectiveness of two non-linear measures, Higuchi's Fractal Dimension (HFD) and Sample Entropy (SampEn), in detecting depressive disorders when applied on EEG. We recorded EEG in 21 participants diagnosed with depressive disorder and 20 healthy age-matched peers. The 10/20 International system for electrode placement was used (19 channels). The HFD and SampEn of EEG signals were used as features for seven machine learning algorithms including Multilayer Perceptron,

Logistic Regression, Support Vector Machines with the linear and polynomial kernel, Decision Tree, Random Forest, and Naïve Bayes classifier, discriminating EEG between healthy control subjects and patients diagnosed with depression. We confirmed earlier observations that both non-linear measures can discriminate EEG signals of patients from healthy control subjects. Besides, our results suggest that if there is a proper feature selection, a useful classification is possible even with a small number of principal components. Average accuracy among classifiers ranged from 90.24% to 97.56%. Among the two measures, SampEn had better performance.

Figure 1: Upper: (left) Principal Component analysis as a feature extraction method in our research; healthy controls are dispersed due to a healthy presence of irregularity in their signals (blue rhombs), and Patients diagnosed with depression are clustering due to the same characteristic (elevated) complexity in their EEG (red stars). Upper right: ROC curve show for our classification. Lower: Leave-one-out classification error (for two out of seven applied models), prior to ten-fold cross-validation we applied in the final version of our work.

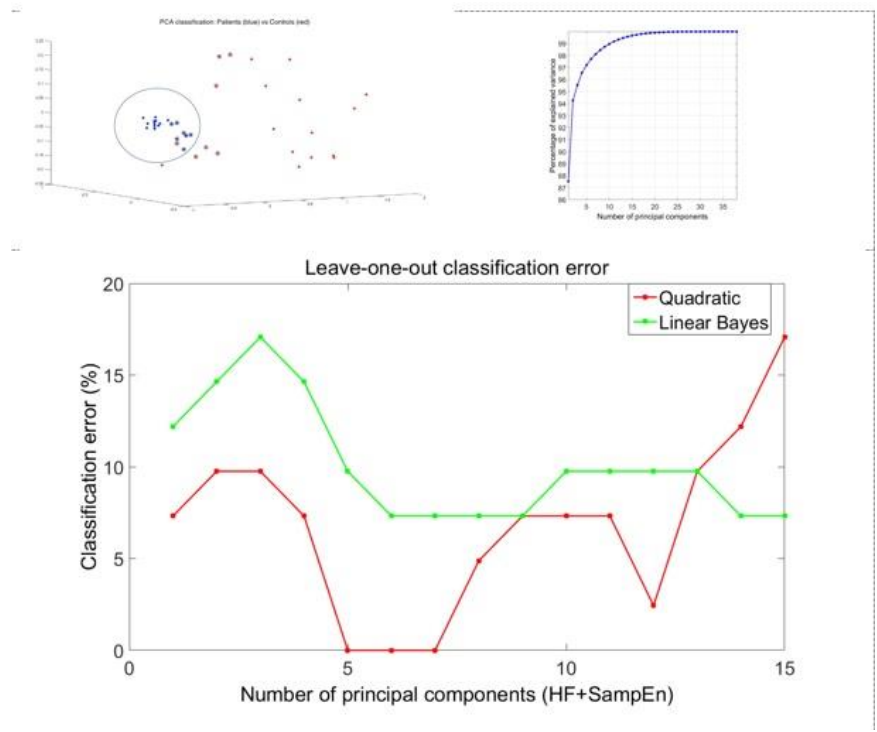

We concluded that using HFD and SampEn and a variety of machine learning methods we can accurately discriminate patients diagnosed with depression vs. controls which can serve as a sensitive, clinically relevant marker of depression. In comparison with previously mentioned studies which also used resting state EEG, we confirmed that the number of electrodes is important because from PCA readings it is clear that every electrode has its own contribution to the result (Čukić et al., 2018). Further, we cannot say that all mentioned studies are giving sufficient information for replication, like for example Bairy, who did not even state what method he used for calculation of fractal dimension, not to mention the algorithm. Others concentrated on improvement of classification but did not undertake all the measures necessary to reach unwarranted optimism in their results (we will discuss it a little bit later). Last but not least all the studies (ours included, although we stated it was a pilot study) had very modest samples, which is affecting the generalizability of the model. Some authors are interpreting their findings somewhat contradictory, for example making a formula (and index) just made of three out of many more measures without the slightest explanation, or the evidence to support that decision.

| Study | Sample | No of electrodes + fs (Hz) | Preprocessing | Features | ML models | Accuracy |
|---|---|---|---|---|---|---|
| Ahmadlou et al 2012. | 12 MDD (+12HC) | 7 256Hz | Wavelets and spectral bands (Fourier), bootstrap | Higuchi's and Katz's FD | Enhanced Probabilistic neural networks | 91.3% |
| Bachman et al, 2018 | 13 MDD+ 13 HC | 1 1KHz | Fourier | HFD, DFA, Lempel-Ziv cmplx and SASI | Logistic regression | 88% |
| Hosseinifard et al, 2015 | 45MDD+ 45HC | 19 1KHz | Standard spectral bands | Power, DFA, Higuci, correlation dimension, Lapynov exp | KNN, LR, linear discriminant | 90% |
| Faust et al, 2014 | 30MDD+ 30HC | 4 (2 left, 2 right) 256Hz | Wavelet package decomposition | ApEn, SampEn, REN, bispectral phase entropy | PNN, SVM, DT, KNN, Naive Bayes, GMM, Fuzzy Gueno Classifier | 99.5% |
| Bairy et al, 2015 | 30D+30HC (left brain only) | ? | Discrete cosine transform | SampEn, FD, CD, Hurst exp, LLE, DFA | DT, KNN, NB, SVM | 93.8% |
| Puthankattil et al, 2012 | 30(16m+14f) +30HC (isti uzorak kao Faust, koautori) | 4 256Hz | Wavelet, Total variation filtering, multiresolution decomposition | Wavelet entropy | RWE, artificial Feed Forward networks | 98,11% |
| Mumtaz et al, 2017 | 33MDD +30HC | 19 (EO, EC) 256Hz | Fourier 10 fold cross validation | Alpha interhemispheric asymmetry | LR, SVM, NB, | 98,4% |
| Cukic et al, 2018. | 26MDD+20HC | 19 1KHz | Broadband EEG, 10 fold cross validation ,PCA | HFD+SampEN | MP, LR, SVM (with linear and polynomial kernel), DT, RF, NB | 97.5% |
| Mohammadi et al., 2015. | 53MDD+43HC | 28 (10/10) 500Hz | Standard bands/FFT LDA, Genetic Algorithm (GA) | spectral | Decision tree (DT) | 80% |
| Acharya et al., 2015. | 15MDD+15HC | 2 left, 2 right, 256Hz | broadband | FD, LLE, SampEn, DFA, H, W-Bx,W,By, EntPh, Ent1,DET, ENTR, LAM, T2 (DDI) | SVM, KNN, Naive Bayes, PNN, DT | 98% |
| Liao et al., 2017. | 12MDD+12HC | 30 500Hz | Common spatial pattern (CSP) | Spectral (CSP) | Kernel-eigen-filter-bank (KEFB-CSB) | 80% |

Table 1: The comparison of above mentioned studies with several characteristics comapared, including their accuracy on classification task. We mentioned here the number of patients and controls, the number of electrodes used for analysis and sampling frequency, whether the signal was analyzed broadband or divided on sub-bands, what methods of machine learning were used and what accuracy the researchers reported.

**Interventional EEG studies**

There are also studies published in the same time interval (2011-2018) based on EEG registration, but the difference from work mentioned above, is they opted to use a stimulus (hence, not resting state EEG), a sound stimulation, or ERP, so we will briefly mention their results here. Cai et al., 2018. Utilized only three electrodes on prefrontal positions to record the signal, while stimulating their participants with a sound. They claim that due to a small number of electrodes which can be easily positioned, their method has a great potential of translation to clinics. Cai used electrophysiological database comprising of 92 depressed patients and 121 healthy controls; resting state EEG was recorded while sound stimulation (they used pervasive prefrontal lobe electrodes on positions Fp1, Fp2, and Fpz). After denoising (Finite Impulse Response, FIR filter) they combined Kalman derivative formula and Discrete Wavelet Transformation, and Adaptive predictor Filter; a total of 270 linear and nonlinear features were extracted (it is not clear what are they). Feature selection was minimal-redundancy-maximal-relevance, which reduced the dimensionality of the feature space. Four classification methods were applied: SVM, KNN, Classification threes and Artificial Neural Networks (ANN). For evaluation, they used 10-fold cross-validation. KNN had the highest accuracy of 79.27%. Their results also show that absolute power of the theta wave might be valid characteristic for predicting the depression. Kalatzis and colleagues published a study about the SVM-based classification system for discriminating depression by using P600 component of ERP signals (Kalatzis et al., 2004). EEG was recorded on 15 electrodes, and a sample comprised of 25 patients and an equal number of healthy controls. The outcomes of SVM classification were selected by Majority vote engine (MVE). Classification accuracy reportedly was 94% when using all leads, and from 92% to 80% when using only right or left points for classification. They concluded that their findings support the hypothesis that depression is associated with the dysfunction of right hemisphere mechanisms mediating the processing of information that assigns a specific response to a particular stimulus. Lee et al., (2011) tried to predict the treatment response of major depressive disorder (given the previous data that initial intent-t-treat response rate is only 50 to 60%). Their study was designed to check whether the connectivity strength of resting state EEG could be a potential biomarker to answer this question. They concluded that '…the stronger the connectivity strengths, the poorer the treatment response.' The experiment also suggested that frontotemporal connectivity strengths could be a potential biomarker to differentiate responders and slow responders or non-responders in MDD. Another two studies from 2013, tried to predict the response of treatment in depression (Olbrich and Arns, 2013; and Arns et al., 2013). The first study evaluated the predictive potentials of several different markers including quantitative EEG (QEEG), connectivity measures, EEG vigilance-based measures, sleep-EEG-related measures and event-related potentials (ERPs). They claimed to review the prognostic potential of mentioned biomarkers for treatment outcomes, but finally focused on standardization of further measurements. A possible reason for this missing conclusion of clear prognostic value could be concentration on standard sub-bands and classical spectral analysis. Even LORETA (together with QEEG) is too complicated to be used in everyday clinical practice, due to limited information it offers (especially for low-density EEG). Another study by Arns claimed that there is no difference between MDD and HC in non-linear EEG measures. According to their reported method, the potential reason for that could be concentration on just one specific band and not on analysis of broadband signal, for many researchers's after (and before) them succeeded to find significant difference by utilizing many nonlinear measures for this kind of detection task (Ahmadlou et al, 2012; Bachmann et al, 2013, 2018; Hosseinifard et al, 2013; Mumtaz et al, 2015; Mohammadi et al, 2015; Čukić et al., 2018). They also claimed that they were 'the first' to use complexity measures in this task. Nandrino and Pezard performed that approach to analysis of EEG in depression in 1994, as well as several other researchers groups (Nandrino and Pezard, 1994). Bachmann and colleagues (2018) applied exactly the same methodology (Lempel-Ziv complexity) and demonstrated significant differentiation between patients and controls. Erguzel and colleagues (Erguzel et al., 2015) tested their optimized classification methods on 147 participants with MDD treated with rTMS. They tested the performance of a genetic algorithm (GA) and a back-propagation neural network (BPNN); they were evaluated using 6-channel pre rTMS EEG patterns of theta and delta frequency bands. By using the reduced feature set, they obtained an increase under the receiver operating curve (AUC) of 0.904. Zhang et al. explored neural complexity in patients with poststroke depression (Zhang et al., 2015) in a resting state EEG study. Their sample comprised of 21 post-stroke patients and 22 ischemic-stroke nondepression and 15 healthy controls: 16 electrodes were used for recording of resting state EEG. Lempel-Ziv complexity (LZC) was used to assess changes in complexity from EEG. PSD (depressed) patients showed lower neural complexity compared with PSND (non-depressed) and Control subjects in the whole brain regions. LZC parameters used for Post-stroke depression (PSD) recognition possessed more than 85% in specificity, sensitivity and accuracy suggesting the feasibility of LZC to serve as screening indicator for PSD. Finally, there were two antidepressive treatment non-response prediction studies: Shahaf et al. (2017), and al-Kaysi et al. (2017). Shahaf and his colleagues developed new electrophysiological attention-associated marker from a single channel (two electrodes: Fpz and one earlobe) using 1 min samples with auditory oddball stimuli and showed to be capable of detecting a treatment-resistant depression (26 patients, 10 controls). Al-Kayasi and team aimed at predicting tDCS treatment outcome of patients with MDD using automated EEG classification. They accurately predicted 8 out of 10 participants when using FC4-AF8 (accuracy 76%), and 10 out of 10 when using CPz-CP2 (accuracy 92%). This finding demonstrates the feasibility of using machine learning to identify patients to respond to tDCS treatment. We are offering the above mentioned studies we review in the following table, which extracted several vital details for the purpose of comparison (although the direct methodological comparison is not possible). Those are the size of the sample, the number of electrodes used, the sampling frequency, linear or nonlinear measures calculated from the raw signal, whether they used sub-bands or the broadband signal for analysis, what machine learning methods they used and finally, what was their obtained accuracy of classification. For this group of studies, it can also be said that they are challenging to compare methodologically, but they are the part of the same effort of showing that not only detection but monitoring and predicting the pace of recovery or output of the treatment is possible.

**On overrated optimism in machine learning and how the present methods can be improved of serve in clinical practice**

To predict clinical outcomes or relapses (for example, after remission in recurrent depression) would be of great clinical significance especially in clinical psychiatry. However, developing a model for predicting a particular clinical outcome for the previously unseen individual have certain challenges both methodological and statistical (Whelan, 2013). A group of authors elucidated risks, pitfalls and recommend the techniques how to improve model reliability and validity in future research (Whelan and Garavan 2013; Gillan and Whelan, 2017; Yahata et al., 2017). The authors described that neuroimaging researchers who start to develop such predictive models are typically unaware of some considerations inevitable to accurately assess model performance and avoid inflated predictions (called 'optimism') (Whelan and Garavan, 2013). They described that in neuroimaging particularly there are several reasons that could lead to misinterpretation of the results of machine learning. Different machine learning methods have been applied to use fRMI (or MRI, or resting state fMRI) data to objectively identify disorder-specific biomarkers for depression. Common characteristic to that kind of research are: classification accuracy is overall 80-90%; the size of sample is typically small to modest (less than 50-100); the samples are usually gathered on a single site, and inter-regional functional connectivity and associated graph-metrics are popular features used for classification (valid for all MRI, fMRI, and rs-fcMRI). Support vector machines (SVM) and its variants are popular but recommendable is the use of embedded regularization frameworks, at least with absolute shrinkage and selection operator (LASO) (Yahata et al., 2017). Leave-one-out and k-fold cross validation are also popular procedures for validation (for model evaluation), and generalization capability of a model is typically untested on an independent sample (Yahata et al., 2017). For model evaluation or even reduction, Vapnik-Chevronenkins dimesion should be used (Vapnik, 1988).

From a methodological point of view there are many problems to resolve. Let's consider them one by one. What is the problem with a generalization? When we test the generalizability, we are basically testing whether or not a classification is effective in an independent (previously not seen/not shown to algorithm before) population. There are several problems with that. If one develop a classifier, on a certain number of learning the data, the classifier can be trained not only on wanted features of the sample but could pick up some general characteristics like gender and age, and seek for that in the test data. Of course, one doesn't want to train the classifier on a general characteristic of a sample; for example, if using nonlinear measures, they can differ because some of the measures change with age (Goldberger et al., 2000) or they can be characteristic for gender (Ahmadi et al., 2013). Also, there is infamous overfitting and the treatment of nuisance variables. Overfitting happens when 'a developed model perfectly describes the entire aspects of the training data (including all underlying relationships and associated noise), resulting in fitting error to asymptotically become zero' (Yahata et al., 2017). For that reason (overfitting) the model will be unable to predict what we want on unseen data (test data). An illustration of overfitting can be seen in the following figure. In neuroimaging (and also in electroencephalographic studies) there are usually many more data points than the number of participants (number of voxels is typical of order 105 for functional and diffusion tensor imaging and of order 106 for structural MR images). The size of the sample is usually small to modest (typically less than 100). Hence, the balancing of the complexity of the model against the sample size is essential for improving prediction accuracy for unseen (test) data (Yahata et al., 2017). How that goal can be achieved? Collect more data. Due to a high cost of all MRI modalities (DTI, FA also), increasing the sample size to order of 103 seem unrealistic for a single laboratory; therefore several collaboration projects are initiated (RDoC, STAR*D, IMAGEN, etc.) where standardized recordings with precisely the same set-up will follow.

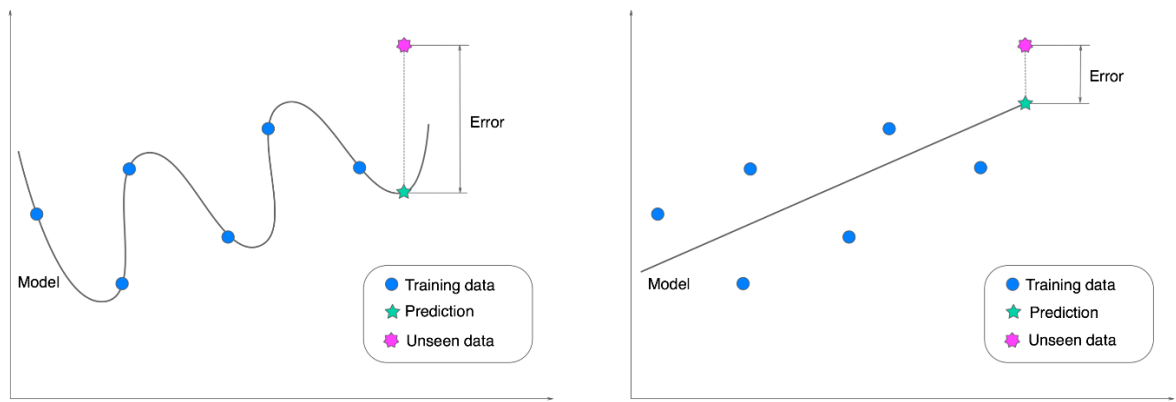

Figure 2: An illustration to overfitting; (a) when a parameter dimension of a model exceeds the sample size of the training data, the model fits not only to the true underlying relationship, but also to noise structure inherent in the data. The ability of the model to predict on unseen data becomes extremely poor. (b) an appropriate procedure to circumvent overfitting is to balance model complexity with the sample size on the training data.

Then with big collaborative efforts, we may approach the conditions to extract genuinely reliable models for clearly defined neuromarkers for future clinical use. Large-scale imaging campaigns and collection of general population data are the condition for translation of those research findings to clinics. By allowing their usual check-up medical data to be a part of such organized collaborative efforts patients would also contribute to the improvement of this precise diagnostic of a (near) future. According to Wessel Kraiij (Data of value, 2017) P4 concept for healthcare improvement stands for: Prediction, Prevention, Personalization and Participation. An important motivation is the observation that healthcare is too focused on disease treatment and not enough on prevention. And another important observation (Kraiij, 2017) is, that treatment and diagnosis are based on population averages. In some cases the treatment have a negative effect. So, there is a lot of room for precision inmporvement (and for other three Ps aside Personalization). In order to do that it is necessary to collect and interpret data of value. The collection, analysis and sharing of the data plays an important role in improvement of the healthcare that we know. In that sense, the first project to implement 4P is SWELL project part of a Dutch national ICT program COMMIT (between 2011 and 2016 in the Netherlands, Leiden University).

Whelan and Garavan (2013) precisely addressed other methodological issues, overfitting included. Their goal was to describe how regression models can appear, incorrectly, to be predictive, and to describe methods for quantifying and improving model reliability and validity.

They explained that '… (regression model)…will result in overfitting and optimism unless particular precaution is taken. Overfitting occurs because a model derived from a particular sample will partly reflect the unique data structure of that particular sample-including the noise in the data' (Whelan and Garavan, 2013). They further elucidate the difference between apparent and actual error leading to unwarranted optimism (this reduction is also known as shrinkage). The solution for this (for predictive models) is to minimize and quantify this inherent optimism. So, quantifying model performance is of extreme significance; there are several methods and techniques to do that. Very useful is the receiver operating curve (ROC) which compares sensitivity versus specificity; it quantifies the model's ability to correctly assign a patient to the disease group. The authors conclude that '…perhaps counterintuitively to those who deal primarily with a general linear model, optimism increases as a function of the decreasing number of participants and the increasing number of predictor variables in the model (model appears better as sample size decreases)' (Whelan and Garavan, 2013).

An illustration of ROC curve is in the Figure below.

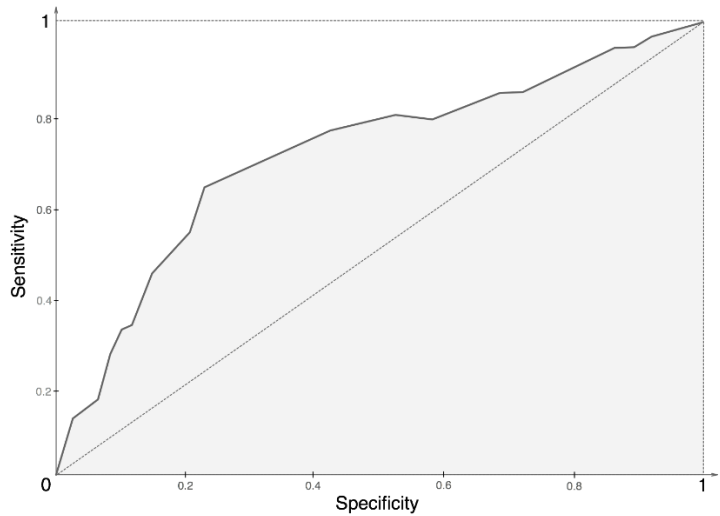

Figure 3: Receiver Operating Characteristic; ROC is displaying sensitivity (correctly identify those with the disease, true positive rate) over specificity (the ability of a test to correctly identify those without the disease, true negative rate). The dashed line marks the random classification accuracy. The shaded area is the so-called area under the curve (AUC) which is a summary metric for the performance of a classifier model.

Despite this expert knowledge about false optimism from our machine learning research, rare is that kind of attempt to quantifying the model performance. What else can be done, when you want to confirm that your model's accuracy is promisingly high? You could collect more data. The theory of data mining is clear; all those models work best on more significant numbers. Therefore collaborative projects in neuroimaging. Use the repository to test your developed model on the unseen cohort. At least, what I learned is that we need statistics to stop making fools from ourselves. Data mining is the art of finding the meaning from supposedly meaningless data. A minimum rate of ten cases per predictor is common (Peduzzi et al., 1996), although not a universal recommendation (Vittinghof and McCulloch, 2007). Optimism can also be lowered with the introduction of the regularization term (Moons et al., 2004). A penalty for a model complexity can constrain the size of parameter values; optimism attenuation should be taken seriously (Zou et al., 2005; Tibshirani et al., 1996). Also, using previous information to constrain model complexity relying on Bayesian approaches is recommendable. Bootstrapping is another helpful method (Efron and Tibshirani, 1993) as well as cross-validation (Efron and Tibshirani, 1997).

Cross-validation tests the ability of the model to generalize and involves separating the data into subsets (10-fold cross-validation is a model that we regularly use, Cukic et al., 2018). Both Kohavi and Ng described the technique (Kohavi et al., 1996; Ng et al., 1997). Beside very useful and efficient ten-fold cross-validation, Elastic Net is useful to optimize parameters. Ng stated that '…optimism becomes unreliable as the probability of overfitting to the test data increases with multiple comparisons' (Ng et al., 1997). One can use several functions available in MATLAB (The MathWorks, Natick, Massachusetts) like lassoglm, bootstrap for bootstrap sampling, or several functions for Bayesian analysis, or crossvalind for testing sets and cross-validation. To conclude Whelan wrote on the importance of keeping training and test subsets completely separate; 'any cross-contamination will result in optimism' (Whelan and Garavan, 2013). When so many brilliant minds focused on solving a problem of reframing nosology in psychiatry, the solution is already visible on the horizon.

Like Paulus illustrated it in his 2015 paper: 'One would like to be able to say, "Mrs. Jones, your depression test has come back and, with the current treatment, you have a 90% chance of being symptom free within the next 6 weeks." However, this statement is still science fiction'.

We hope, not for so long.